# Emergence of Bending Power Law in Higher-Order Networks

Zhaohua Guo [1], Yuan Yuan[2], Rui Miao[1]**, Jin-Li Guo [2]**, Jeffrey Forrest[3]

*1.School of Naval Architecture, Ocean & Civil Engineering, Shanghai Jiao Tong University, 200240, China ;*

*2. Business School, University of Shanghai for Science and Technology, Shanghai, 200093, China*

*3. Department of Accounting Economics Finance, Slippery Rock University, PA 16057, USA*

## Abstract

In the past two decades, a series of important results have been established in the empirical and theoretical modeling of complex networks, although considered are mainly pairwise networks. However, with the development of science and technology, an increasing number of higher-order networks with many-body interactions have gradually moved to the center stage of research when real-life systems are investigated. In the paper, the concept of higher-order degree is introduced to higher-order networks, and a bending power law (BPL) model with continuous-time growth is proposed. The evolution mechanism and topological properties of the general higher-order network are studied. The batch effect of low dimensional simplex is considered. The model is analyzed by using the mean-field method and Poisson process theory. The stationary average higher-order degree distribution of simplices is expressed analytically. The obtained analytical results agree well with those observed through simulations. In particular, this paper shows that the higher-order degree distribution of simplices in the network processes a property of bending power law, and the scale-free property of the higher-order degree is controlled by the higher-order edge, the simplex dimension and the feature parameter of the model. The BPL model of higher-order networks not only generalizes the NGF model, but also the famous scale-free model of complex networks to higher-order networks.

---

** Corresponding author. Email: miaorui@sjtu.edu.cn
** Corresponding author. Email: phd5816@163.com

## 1. Introduction

The emergence of certain characteristics in complex systems depends largely on the way that the basic components interact with each other. The core of complexity science is the focus on the concept of emergence. Emergence in complex systems has been a hot topic of research for scientists. Complex networks are a powerful tool for describing complex systems; in the past two decades, we have seen the birth and development of the interdisciplinary field of network science [1-7]. The study of complex networks has been a highly active field for over two decades attracting researchers from different fields, such as physics, computer science, graph theory, transportation, and economics. The fields involved in the study range from internet networks to WeChat social networks, from power networks to transportation networks, from scientific cooperation networks to various economic, political, and social relationship networks. For twenty years, scientists have constructed a large number of network models, the core content of which is the study of the emergence of scaling laws in networks.

Complex networks study such networks that are formed by the connection of a pair of nodes. However, with the development of society and the advancement of technology, the complexity of systems considered has been increasing, and the established theory of complex networks is no longer able to adequately describe real-life systems, such as, the competition network among three or more species that fight for food or territory in an ecosystem, brain structure network, protein interaction network, semantic network, and cooperation network of scientists [8,9]. With the advent of the big data era, there is a growing emergence of higher-order networks that can be more appropriately used to characterize multiple interactions. In recent years, the study on higher-order networks has gradually gained attention [8-14]. Higher-order networks include hypergraph-based and simplex-based networks [8,13].

In terms of dynamic evolution models of hypergraph-based networks, the growth of a higher-order network may involve batch addition of nodes. Wang et al. proposed an evolutionary model based on hypergraphs that generates a higher-order network through using a growth and preferential attachment mechanism. In this model, each time $m_1$ new nodes are added, these nodes combine with one existing node in the network to form a hyperedge, and only one hyperedge is added at each time step [15]. Hu et al. proposed another evolutionary model based on hypergraphs, which uses the same preferential attachment mechanism as the model by Wang et al. Each time a new node is added, it is combined with many existing nodes in the network to form a hyperedge, and only one hyperedge is added at each time step [16]. Both of these models only add one hyperedge at each time step, so they cannot degenerate into the BA model as studied in complex networks [2]. Guo and Zhu proposed a unified model to study the emergence of scaling laws in higher-order networks based on hypergraphs. In this model, $m_1$ new nodes are added at each timestep, and combined with $m_2$ nodes in the existing network to form a hyperedge, and $m$ new hyperedges are added at each timestep [17]. These models describe the evolution of high - order networks based on hypergraphs well, but investigate only the statistical properties of nodes on hyperedges. Although higher-order networks based on simplices are abstract, algebraic topology tools can be used to characterize the higher-order structure of such networks. Bianconi et al. introduced a concept of generalized degrees in simplicial complexes [13], and employed algebraic topology tools to investigate high-order networks. They proposed a network geometry with flavor (NGF) model. It is an important evolution model of simplicial complexes [13]. The NGF model generates simplicial complexes of any dimension $d$ and have very distinct statistical properties. The NGF model generates only one $d$-dimensional simplex at each time step. Kovalenko et al. generalized the NGF model to the case where multiple 2-dimensional simplices are created at each time step [9]. The main goal of this paper is to propose a bending power law (BPL) model on simplicial complexes, which extends the BA model to higher-order networks and explores the emergence of bending power law in higher-order networks.

The rest of this study is structured as follows. In the next section, we introduce the concept of higher-order networks based on hypergraphs and simplicial complexes, and the concept of the higher-order degree. In the third section, we introduce two basic models in higher-order networks and propose a bending power law (BPL) model of simplicial complexes. In the fourth section, batch effects based on low-dimensional simplex when a new node arrives, we use Poisson process theory and a mean-field method to analyze the BPL model. We obtain an analytical expression for the stationary average higher-order degree distribution of this network and simulate the model. Both of these efforts show the emergence of scaling laws in higher-order networks based on simplices. The last section summarizes what is achieved in this paper.

## 2. The concept of higher-order networks

Hypernetworks are divided into two types: higher-order networks based on hypergraphes and higher-order networks based on simplices. Let $U$ be a set, $V \subset U$. We call U a set of nodes. The mathematical definition of a hypergraph is given as follows: A hypergraph $H$ is equal to the following ordered pair $H = (V, E^h)$, where $V = \{v_1, v_2, \cdots, v_n\} \subset U$, $I = \{1, 2, \dots, m\}$ stands for an index set, and $E^h = \{E_i\}_{i \in I}$ such that for any $i \in I$, $\Phi \neq E_i \subset V$ and $\cup_{i=1}^{m} E_i = V$. Each $E_i$ is called as a hyperedge.

Two nodes in a hypergraph are adjacent if there is a hyperedge which contains both nodes. In particular, if $\{v\}$ is an hyperedge then $v$ is adjacent to itself. Two hyperedges in a hypergraph are incident if their intersection is not empty. The order of the hypergraph H $= (V, E^h)$ is the cardinality of V, i.e. $|V|$; the size of $E^h$ is the cardinality of $E^h$, i.e. $|E^h| = m$. If $|V|$ and $|E^h|$ are finite, $H$ is called a finite hypergraph. If $|E_i| = c (i = 1, 2, \cdots, m)$, it is called a uniform hypergraph. If $|E_i| = 2 (i = 1, 2, \cdots, m)$, the hypergraph $H = (V, E^h)$ degenerates into a graph.

With the mathematical definition of a hypergraph in place, we can define a higher-order network based on hypergraphs. Suppose that $\Omega = \{(V, E^h) | V \subset U, (V, E^h)$ is a finite hypergraph$\}$, $G$ is a map from $[0, +\infty)$ to $\Omega$, and $N^*(t)$ denotes the total number of times that the hypergraphs have been changed by time $t$. If $\{N^*(t), t \geq 0\}$ is a stochastic process, for sufficiently large time $t$, we call $G(t)$ a higher-order network. The hyperdegree of a node $i$ is defined as the number of hyperedges containing the node, denoted as $k_i^h$.

Let $\alpha = \{v_{\alpha,0}, v_{\alpha,1}, \cdots, v_{\alpha,d-1}, v_{\alpha,d}\}$ be a subset of $V = \{v_1, v_2, \cdots, v_n\} \subset U$, and $\alpha \neq \Phi$, $\alpha$ is called a $d$-dimensional simplex [13]. A face of a $d$-dimensional simplex $\alpha$ is a simplex $\alpha'$ formed by a proper subset of nodes of the simplex , i.e. $\alpha' \subset \alpha$. A simplicial complex $K$ based on $V = \{v_1, v_2, \cdots, v_n\}$ is a set of simplices that satisfy the following conditions:

(a) if a simplex $\alpha$ belongs to the simplicial complex $K$, then any face $\alpha'$ of the simplex $\alpha$ is also included in the simplicial complex $K$;

(b) given two simplices of the simplicial complex $\alpha \in K$ and $\beta \in K$ then either their intersection belongs to the simplicial complex, i.e. $\alpha \cap \beta \in K$, or their intersection is null, i.e. $\alpha \cap \beta = \Phi$.

The largest dimension of the simplices in $K$ is called the dimension of simplicial complex K, denoted as $\dim K$. If K is a simplicial complex, a $r$-skeleton of $K$ is formed by a set of simplices that their dimension does not exceed $r$.

The higher-order network based on simplices defined as follows: Let $\Omega = \{(V,K)|V \subset U, \ K$ is a finite dimensional simplicial complex based on V$\}$, $G$ be a map from $[0, +\infty)$ to $\Omega$, and $N^*(t)$ denote the total number of times that the simplicial complexes have been changed by time $t$. If $\{N^*(t), t \geq 0\}$ is a stochastic process, we call $G(t)$ a higher-order network based on simplices for sufficiently large time $t$. The $d$-order degree $h_{d,e}^{\alpha}$ of an $e$-dimensional simplex $\alpha$ indicates the number of $d$-dimensional simplices incident to the simplex $\alpha$.

Let $d(t)$ be the dimensions of $G(t) = \left(V(t), \ K(t)\right)$, If $d(t) \equiv 1$ holds, the higher-order network based on simplices degenerates into a complex network, which shows that higher-order networks based on simplices are a generalization of the concept of complex networks.

## 3. Model Description

### 3.1. Evolution model of higher-order networks based on hypergraphs

Barabási and Albert proposed the BA model to describe network evolutions that is useful for explaining the ubiquitous phenomenon of a power-law distribution in complex networks [2]. Their work indicates that the growth and preferential attachment play an important role in network development. In the BA model, the preferred probability is a homogeneous linear function. Guo and Zhu generalized the BA model to higher-order networks based on hypergraphs [17]. This new model refers to the higher-order network that satisfy the following two rules: (1) *Growth*: Starting with a small number $m_0$ of nodes and a hyperedge containing these $m_0$ nodes. The arrival process of nodes follows a Poisson process with rate $\lambda$. At time $t$, when a batch of $m_1$ new nodes is added to the network, these $m_1$ new nodes and $m_2(\leq m_0)$ previously existing nodes are encircled by a new hyperedge, $m(\leq m_0)$ new hyperedges are constructed with no repetitive hyperedges; (2) *Preferential* *attachment*: when selecting an existing node $i$ in the network to connect to a new node, the selection probability $W$ of node $i$ depends on the hyperdegree $k_i^h$ of node $i$, satisfying:

$$W(k_i^h) = \frac{k_i^h}{\sum_j k_j^h}. \tag{1}$$

Equation (1) represents the preferential attachment of new nodes to old nodes based on the homogeneous linear function of the degrees of the old nodes. When $m_1 = m_2 = 1$, this model degenerates into the BA model developed for complex networks.

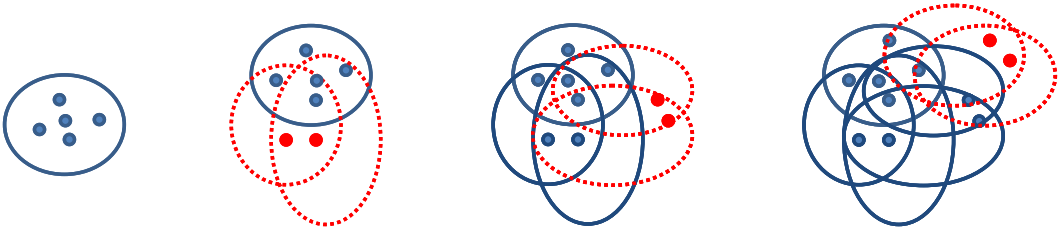

**Fig. 1.** Schematic diagram of the first three steps of evolution of the higher-order network model based on hypergraph $(m_0 = 5, m_1 = 2, m_2 = 3, m = 2)$

### 3.2. NGF model of higher-order networks based on simplices

The evolution model of higher-order networks based on hypergraphs only studied the power law feature of the hyperdegree distribution. In higher-order networks based on simplices, there are the $d$-order degree of nodes (0-dimensional simplexes) and $e(e > 0)$-dimensional simplexes. In order to investigate the topological properties of higher-order networks based on simplexes, Bianconi et al. established the NGF model, which refers to a higher-order network satisfying the following two rules: (1) *Growth*: At time $t = 1$, the network is formed a single $d$-dimensional simplex. At every timestep a new $d$-dimensional simplex formed by one new node and existing $(d-1)$-dimensional face is added to the simplicial complex; (2) *Preferential attachment*: The probability $W$ that the new $d$-dimensional simplex is glued to a $(d-1)$-dimensional face $\alpha$ depends on the $d$-order degree of the face $\alpha$ and the flavor $s \in \{-1,0,1\}$, and given by

$$W(n_\alpha) = \frac{1 + sn_\alpha}{\sum_\alpha (1 + sn_\alpha)} \quad , \tag{2}$$

where $n_\alpha = h_{d,d-1}^\alpha - 1$.

When $s = 1$, it reflects the probability that the new $d$-dimensional simplex is glued to a $(d-1)$-dimensional face $\alpha$ depends on the homogeneous linear function of $d$-order of degree of the $(d-1)$-dimensional face. In this case, the NGF model degenerates into the BA model with only one new edge ($m = 1$), while the BA model adds $m$ new edges at each timestep.

### 3.3. BPL model of higher-order networks based on simplices

Although the NGF model can describe the evolution of a class of higher-order networks well, its flavor parameter $s$ only take one of three possible values $s \in \{-1, 0, 1\}$. We propose a bending power law (BPL) model of higher-order networks. To this end, let us define the evolution process of the BPL model as follows：

(1) *Growth*: Starting with $m_0$ $d$-dimensional simplices in a simplicial complex, the arrival of nodes follows a Poisson process with rate $\lambda$. At time $t$, when one new node arrives, $m$ new $d$-dimensional simplices formed by the new node and previously existing $(d-1)$-dimensional face are added to the network; and no repeating simplex is allowed；(2) *Preferential attachment*: The probability $\coprod$ that the new $d$-dimensional simplex is glued to a $(d-1)$-dimensional face $\alpha$ depends on the $d$-order degree of face $\alpha$ and feature parameter $p \in [-1,1]$, is given by

$$\coprod(h_{d,d-1}^\alpha) = \frac{ph_{d,d-1}^\alpha + q}{\sum_{\beta \in S_{d-1}(t)}(ph_{d,d-1}^\beta + q)} \quad , \tag{3}$$

where $p + q = 1$; $S_{d-1}(t)$ denotes the set of $(d-1)$- dimensional simplices in the network at time $t$.

If $q \neq 0$, the preferential attachment probability is a non-homogeneous linear function of $h_{d,d-1}^\alpha$. If $p \in \{-1,0,1\}$ and $m = 1$, this model degenerates into the NGF model. If $p = 1$ , $d = 1$, this model degenerates into the BA model.

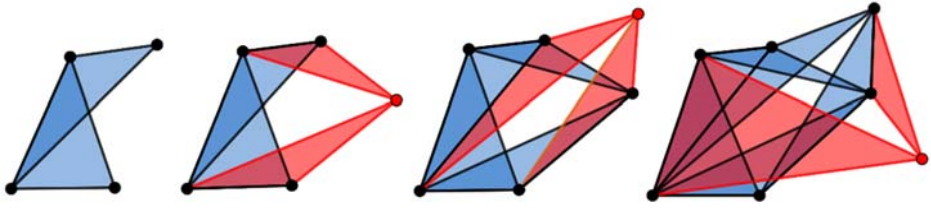

**Fig. 2.** Schematic illustration of the first three steps of the evolution for the model of the higher-order networks based on simplices ( $m_0 = 2, d = 2, m = 2$).

## 4. Model analysis and simulation

Let

$$N(t) = \{Number\ of\ nodes\ in\ the\ network\ at\ time\ t\} - m_0(d+1).$$

As the arrival process $N(t)$ of nodes is a Poisson process with rate $\lambda$, according to the theory of Poisson processes, $E[N(t)] = \lambda t$. The time dependence of the connectivity of a given simplex can be calculated analytically using a mean-field approach.

Let $t_\alpha$ denote the time when the last node of simplex $\alpha$ enters the network, $h_{d,d-1}^\alpha(t)$ the $d$-order degree of the ($d$-1)-dimensional simplex $\alpha$ at time $t$. Assume that $h_{d,d-1}^\alpha(t)$ is a continuous real-valued variable, and since the rate of change of $h_{d,d-1}^\alpha(t)$ is proportional to the probability $\prod(h_{d,d-1}^\alpha(t))$, $h_{d,d-1}^\alpha(t)$ satisfies the following equation.

$$\frac{\partial h_{d,d-1}^\alpha}{\partial t} = m\,\lambda\,d\,\frac{ph_{d,d-1}^\alpha + q}{\sum_\beta (ph_{d,d-1}^\beta + q)}. \tag{4}$$

Factor $m$ on the right-hand side of the equation is due to the reason that simplex $\alpha$ may be selected in the first, second, third, ..., $m$th step. Factor $d$ is due to the production of $d$ new ($d-1$)-dimensional simplices each time, so the arrival of new ($d-1$)-dimensional simplices in the system follows a Poisson process with rate $\lambda d$.

The second sum in the denominator of the equation (4) represents $q$ times the total number of ($d$-1)-dimensional faces in the simplicial complex at time $t$. At each time step, we add $m$ new $d$-dimensional simplices, each of which contributes $d$ new ($d$-1)-dimensional faces (ignoring any overlap between new simplices).

Therefore, for $t \gg 1$, the approximate number of $(d\text{-}1)$-dimensional faces is:

$$\sum_{\alpha \in S_{d-1}(t)} 1 \approx c_{m,1}(d-1)d\lambda t \,,$$

where $c_{m,1}(e) = \begin{cases} 1, & \text{if } e = 0 \\ m, & \text{if } e > 0 \end{cases}$.

For $t \gg 1$, without considering the overlapping new face, in the first sum in the denominator of the equation (4), one new $(d-1)$-dimensional face contributes to one $d$-order degree, and the $d$-order degree of the selected $(d-1)$-dimensional face also increases by one. Therefore,

$$\sum_{\beta \in S_{d-1}(t)} h_{d,d-1}^{\beta} \approx m(d+1)\lambda t \,.$$

Thus, we obtain that:

$$\sum_{\beta \in S_{d-1}(t)} (p h_{d,d-1}^{\beta} + q) = (p(d+1) + qd \frac{c_{m,1}(d-1)}{m})m\lambda t \,. \tag{5}$$

Substituting Equation (5) into Equation (4), we have

$$\frac{\partial h_{d,d-1}^{\alpha}}{\partial t} = \frac{d\,(\,p h_{d,d-1}^{\alpha} + q\,)}{\left(\,(q \dfrac{c_{m,1}(d-1)}{m} + p)d + p\,\right)\,t} \,. \tag{6}$$

For $e < d\text{-}1$, the probability of increasing the $d$-order of $e$-dimensional simplex $\delta$ is approximately

$$\begin{aligned}
\amalg_e(\delta) &= \sum_{\alpha \in S_{d-1}|\alpha \supseteq \delta} \amalg(h_{d,d-1}^{\alpha}) \\
&= \frac{1}{\sum\limits_{\beta \in S_{d-1}(t)}(p h_{d,d-1}^{\beta} + q)} \sum_{\alpha \in S_{d-1}|\alpha \supseteq \delta}(p h_{d,d-1}^{\alpha} + q)
\end{aligned} \tag{7}$$

$$\sum_{\alpha \in S_{d-1}|\alpha \supseteq \delta}(p h_{d,d-1}^{\alpha} + q) = p \sum_{\alpha \in S_{d-1}|\alpha \supseteq \delta} h_{d,d-1}^{\alpha} + q \sum_{\alpha \in S_{d-1}|\alpha \supseteq \delta} 1 \,. \tag{8}$$

From reference [10], it follows that for $e_1 > e$,

$$h_{d,e}^{\beta} = \frac{1}{\dbinom{d-e}{e_1-e}} \sum_{\alpha \in S_{e_1}|\alpha \supseteq \beta} h_{d,e_1}^{\alpha} \,. \tag{9}$$

From Equation (9)，we have

$$\sum_{\alpha \in S_{d-1}|\alpha \supseteq \delta} h_{d,d-1}^{\alpha} = (d-e)h_{d,e}^{\delta},$$  (10)

where $h_{d,e}^{\delta}(t)$ represents the d-order degree of the e-dimensional simplex $\delta$ at time t. According to reference [18], we have

$$\sum_{\alpha \in S_{d-1}|\alpha \supseteq \delta} 1 = h_{d-1,e}^{\delta} = \begin{cases} m + (d-1)h_{d,e}^{\delta}, & if \ e = 0 \\ 1 + (d-e-1)h_{d,e}^{\delta}, & if \ e > 0, \end{cases}$$

$$= c_{1,m}(e) + (d-e-1)h_{d,e}^{\delta}$$  (11)

where $c_{1,m}(e) = \begin{cases} m, & \text{当} e = 0 \\ 1, & \text{当} e > 0 \end{cases}$

Substituting Equations (11) and (12) into Equation (8), we obtain:

$$\sum_{\alpha \in S_{d-1}|\alpha \supseteq \delta} (ph_{d,d-1}^{\alpha} + q) = qc_{1,m}(e) + (d-e-q)h_{d,e}^{\delta}.$$  (12)

At each timestep, one 0-dimensional simplex is generated, while $d \ e(e > 0)$-dimensional simplices are generated. New $e$-dimensional simplices arrive in the system according to a Poisson process with rate $\lambda c_{d,1}(e)$. Assume that $h_{d,e}^{\delta}(t)$ is a continuous real-valued variable. Since the rate of change of $h_{d,e}^{\delta}(t)$ is proportional to the probability $\coprod_e(\delta)$, we know that $h_{d,e}^{\delta}(t)$ satisfies the following dynamic equation

$$\frac{\partial h_{d,e}^{\delta}}{\partial t} = mc_{d,1}(e)\lambda \frac{1}{\sum_{\beta \in S_{d-1}(t)}(ph_{d,d-1}^{\beta} + q)} \sum_{\alpha \in S_{d-1}|\alpha \supseteq \delta}(ph_{d,d-1}^{\alpha} + q).$$  (13)

Substituting Equations (5) and (12) into Equation (13) leads to

$$\frac{\partial h_{d,e}^{\delta}}{\partial t} = c_{d,1}(e) \frac{qc_{1,m}(e) + [(d-e)-q]h_{d,e}^{\delta}}{(p(d+1) + qd \frac{c_{m,1}(d-1)}{m})t}.$$  (14)

By neglecting the calculation of connections to adjacent faces, solving Equation (14) yields.

$$h_{d,e}^{\delta}(t) = \begin{cases} \dfrac{d-e}{d-e-q}c_{1,m}(e)\left(\dfrac{t}{t_{\delta}}\right)^{\frac{(d-e-q)c_{d,1}(e)}{dc_{m,1}(d-1)\frac{q}{m}+p(d+1)}} - \dfrac{qc_{1,m}(e)}{d-e-q}, & if \ e \neq d-q \\ \dfrac{qc_{d,1}(e)c_{1,m}(e)}{dc_{m,1}(d-1)\frac{q}{m}+p(d+1)}\ln\dfrac{t}{t_{\delta}}+c_{1,m}(e), & if \ e = d-q \end{cases} \tag{15}$$

Case 1, $p \geq 0$.

If $e \neq d-q$, we have from Equation (15)

$$P(h_{d,e}^{\delta}(t) \geq k) = P(t_{\delta} \leq (\dfrac{(d-e)c_{1,m}(e)}{((d-e-q)k+qc_{1,m}(e)})^{\frac{dc_{m,1}(d-1)\frac{q}{m}+p(d+1)}{(d-e-q)c_{d,1}(e)}}t). \tag{16}$$

Assume that node $i$ enters the network at time $t_{\delta}$, that is, $t_{\delta} = t_i$. Then, we have

$$\gamma = \dfrac{dc_{m,1}(d-1)\dfrac{q}{m}+p(d+1)}{(d-e-q)c_{d,1}(e)}, \tag{17}$$

$$B = (d-e)c_{1,m}(e),$$

$$A = d-e-q,$$

By Poisson process theory [19], we know that the arrival time $t_i$ of node $i$ follows a Gamma distribution $\Gamma(i,\lambda)$ with parameters $i$ and $\lambda$

$$P(t_i \leq x) = 1 - e^{-\lambda x}\sum_{l=0}^{i-1}\dfrac{(\lambda x)^l}{l!}.$$

Therefore,

$$P(t_i \leq (\dfrac{B}{Ak+qc_{1,m}(e)})^{\gamma}t) = 1 - e^{-\lambda t\ (\frac{B}{Ak-qc_{1,m}(e)})^{\gamma}}\sum_{l=0}^{i-1}\dfrac{1}{i!}(\lambda t\ (\dfrac{B}{Ak+qc_{1,m}(e)})^{\gamma})^i. \tag{18}$$

Substituting Equation (18) into Equation (16), we have

$$P(h_{d,e}^{\delta}(t) \geq k) = 1 - e^{-\lambda t\ (\frac{B}{Ak+qc_{1,m}(e)})^{\gamma}}\sum_{l=0}^{i-1}\dfrac{1}{i!}(\lambda t\ (\dfrac{B}{Ak+qc_{1,m}(e)})^{\gamma})^i. \tag{19}$$

The following results prove that there exists a stationary average distribution of the $d$-degrees of $e$-

dimensional simplices in the network. From Equation (19), we have

$$P(h_{d,e}^{\delta}(t)=k) \approx \frac{\partial P(h_{d,e}^{\delta}(t)<k)}{\partial k}$$

$$=\gamma A(\frac{B}{Ak+qc_{1,m}(e)})^{\gamma+1}\frac{\lambda t}{B}\frac{(\lambda t(\frac{B}{Ak+qc_{1,m}(e)})^{\gamma})^{i-1}}{(i-1)!}e^{-\lambda t(\frac{B}{Ak+qc_{1,m}(e)})^{\gamma}} \tag{20}$$

From Equation (20), the stationary average distribution of the $d$-degrees of $e$-dimensional simplices in the non-homogeneous linear preferential attachment simplicial complex network is given by

$$P_{d,e}(k) \approx \lim_{t \to \infty}\frac{1}{\sum\limits_{\alpha \in S_e(t)}1}\sum\limits_{i}c_{m,1}(e)c_{d,1}(e)P(h_{d,e}^{\delta}(t)=k)$$

When one new node arrives, we add $m$ new d-dimensional simplices, each contributing $d$ new $e$-faces (ignoring any overlap between new simplices, $0 < e < d$). Therefore, for $t \gg 1,$ the number of $e(<d)$-faces is approximately given by

$$\sum\limits_{\alpha \in S_e(t)}1 \approx c_{m,1}(e)c_{d,1}(e)\lambda t . \tag{21}$$

Hence,

$$P_{d,e}(k) \approx \gamma \frac{A}{B}(\frac{B}{Ak+qc_{1,m}(e)})^{\gamma+1}$$

Thus, the d-degree of $e(<d)$-dimensional simplices $\delta$ is given by

$$P_{d,e}(k) = \gamma \frac{d-e-q}{(d-e)c_{1,m}(e)}(\frac{(d-e)c_{1,m}(e)}{(d-e-q)k+qc_{1,m}(e)})^{\gamma+1} . \tag{22}$$

Equation (22) indicates that the $d$-degree of $e(<d)$-dimensional simplices $\delta$ in the simplicial complex network exhibits the scale-free property controlled by characteristic parameter $p$, dimensions $d$ and $e$. The power-law exponent of the higher-degree distribution tail is $\gamma+1$.

If $e = d-q$, we have

$$P(h_{d,e}^{\delta}(t) \geq k) = 1-e^{-\lambda te^{-\frac{q}{m}\frac{dc_{m,1}(d-1)+p(d+1)}{qc_{1,m}(e)c_{d,1}(e)}(k-c_{1,m}(e))}}\sum\limits_{l=0}^{i-1}\frac{1}{i!}(\lambda te^{-\frac{q}{m}\frac{dc_{m,1}(d-1)+p(d+1)}{qc_{1,m}(e)c_{d,1}(e)}(k-c_{1,m}(e))})^{i} . \tag{23}$$

The following results prove that there exists a stationary average distribution of the $d$-degree of $e$-dimensional simplices. From Equation (23), we have

$$P(h_{d,e}^{\delta}(t)=k) \approx \frac{\partial P(h_{d,e}^{\delta}(t)<k)}{\partial k}$$

$$= \lambda t \frac{q}{m} \frac{dc_{m,1}(d-1)+p(d+1)}{qc_{1,m}(e)c_{d,1}(e)} e^{-\frac{q}{m}\frac{dc_{m,1}(d-1)+p(d+1)}{qc_{1,m}(e)c_{d,1}(e)}(k-c_{1m}(e))} \frac{\left(\lambda te^{-\frac{q}{m}\frac{dc_{m,1}(d-1)+p(d+1)}{qc_{1,m}(e)c_{d,1}(e)}(k-c_{1m}(e))}\right)^{i-1}}{(i-1)!} e^{-\lambda te^{-\frac{q}{m}\frac{dc_{m,1}(d-1)+p(d+1)}{qc_{1,m}(e)c_{d,1}(e)}(k-c_{1m}(e))}} \quad . \quad (24)$$

According to Equation (24), the stationary average distribution of the $d$-degree of $e$-dimensional simplices in the network is given by

$$P_{d,e}(k) \approx \frac{qdc_{m,1}(d-1)+pm(d+1)}{qmc_{1,m}(e)c_{d,1}(e)} e^{-\frac{q}{m}\frac{dc_{m,1}(d-1)+p(d+1)}{qc_{1,m}(e)c_{d,1}(e)}(k-c_{1m}(e))} \quad , \quad (25)$$

Equation (25) indicates that this type of the network does not have any scale-free property, and the $d$-degree of $e$-dimensional simplices follows an exponential distribution.

It can be proved that, when $d-e-q \rightarrow 0$, Equation (22)$\rightarrow$Equation (25). Therefore, equation (22) is referred to as a power-law distribution of bending.

Case 2，$p<0$

If $e=d-1$,

When $d>1$，from Equation (6), we have

$$\frac{\partial h_{d,d-1}^{\alpha}}{\partial t} = d \frac{ph_{d,d-1}^{\alpha}+q}{(d+p)t} \quad , \quad (26)$$

and $h_{d,d-1}^{\alpha}(t_{\alpha})=1$. So, we obtain the solution of Equation (26) as follows:

$$h_{d,d-1}^{\alpha}(t) = \frac{1}{p}(\frac{t_{\alpha}}{t})^{\frac{-pd}{d+p}}+1-\frac{1}{p}$$

If $k \leq [1-\frac{1}{p}]$，then

$$P(h_{d,d-1}^{\alpha}(t) \geq k) = P(t_{\alpha} \leq (pk-p+1)^{\frac{d+p}{pd}}t)$$

Let $t_\alpha = t_i$. Then, we have

$$P(h_{d,d-1}^\alpha(t) \geq k) = 1 - e^{-\lambda(pk-p+1)^{-\frac{d+p}{pd}} t} \sum_{l=0}^{i-1} \frac{(\lambda(pk-p+1)^{-\frac{d+p}{pd}} t)^l}{l!}$$

$$P(h_{d,d-1}^\alpha(t) = k) \approx \frac{d+p}{pd} \lambda t p (pk-p+1)^{\frac{d+p}{pd}-1} \frac{(\lambda t(pk-p+1)^{\frac{d+p}{pd}})^{i-1}}{(i-1)!} e^{-\lambda t(pk-p+1)^{\frac{d+p}{pd}}},$$

Therefore,

$$P_{d,d-1}(k) \approx \frac{d+p}{d}(pk-p+1)^{-\frac{d+p}{pd}-1}, \quad k = 1,2,\ldots,[1-\frac{1}{p}], \tag{27}$$

Since $p \in [-1,0)$ and $d > 1$, Equation (27) means that $k$ can only take the value 1 or 2. From Equation (27), we have $P_{d,d-1}(1) = 1 + \frac{p}{d}$; and therefore,

$$P_{d,d-1}(k) = \begin{cases} 1 + \frac{p}{d}, & if \ k = 1 \\ -\frac{p}{d}, & if \ k = 2 \end{cases}. \tag{28}$$

If $0 \leq e < d-1$, because d-e is an integer, it means that $d-e \geq 2$. Since $q \in [0,2]$, therefore, $d-e-q \geq 0$. It can be seen from Equation (12) that

$$\sum_{\alpha \in S_{d-1}|\alpha \supseteq \delta} (ph_{d,d-1}^\alpha + q) = p(d-e)h_{d,e}^\delta + q = qc_{1,m}(e) + (d-e-q)h_{d,e}^\delta > 0.$$

The distribution of the $d$-degree of $e$-dimensional simplices follows Equation (22) or Equation (25).

We conducted computer simulations of the model for $d$=2, $m$=2, and network size N=500. Numerical simulations indicated that this network evolves into a scale invariant state that each $e$-dimensional face has $k$ $d$-dimensional simplices following a power-law distribution of bending (see Fig. 4-Fig.6). This distribution is heavy-tailed. All numerical results agree well with this prediction. From Fig.4, it follows that as $p \to 0^+$, the distribution of the 2-order degree of 1-dimensional simplices in the model is bent. The closer $p$ is to $0^+$, the more curved the distribution is. As we can see from Fig.4-Fig.6, the scale-free property of the higher-order degree of the higher-dimensional simplex is greatly affected by the feature parameter of the model, while that of the higher-order degree of the lower-dimensional simplex is less affected by the feature parameters.

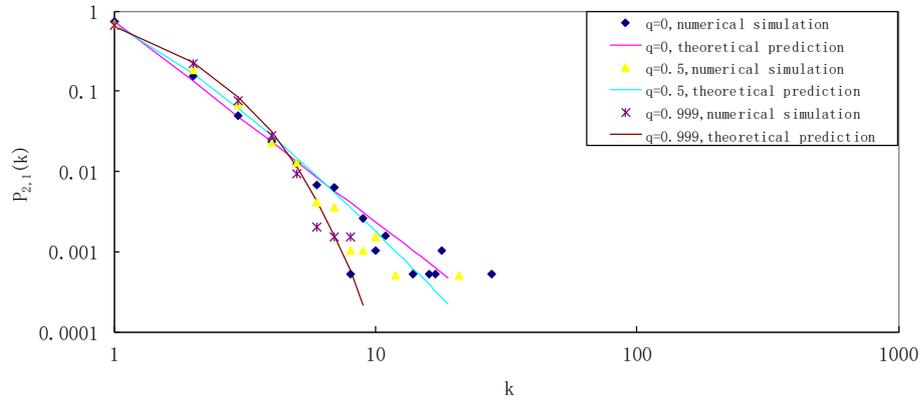

**Fig. 4. The distribution of the 2-order degree of 1-dimensional simplices in the model ( $p = 1 - q \geq 0$ ).** The figure is in log–log scale. ◆ denotes the numerical simulation of the BPL model as $q$=0; — denotes the theoretical prediction of the BPL model as $q$=0; ▲ denotes the numerical simulation of the BPL model as $q$=0.5; — denotes the theoretical prediction of the BPL model as $q$=0.5; * denotes the numerical simulation of the BPL model as $q$=0.999; — denotes the theoretical prediction of the BPL model as $q$=0.999.

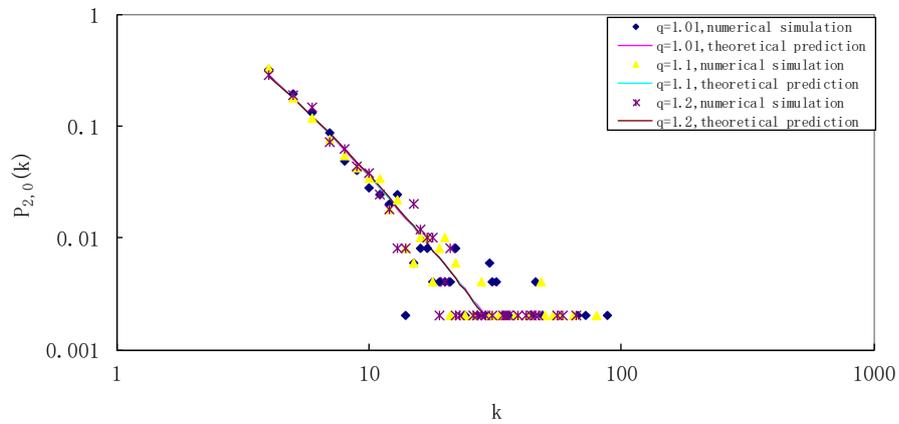

**Fig. 5. The distribution of the 2-order degree of 0-dimensional simplices in the model ( $p = 1 - q < 0$ ).** The figure is in log–log scale. ◆ denotes the numerical simulation of the BPL model as $q$=1.01; — denotes the theoretical prediction of the BPL model as $q$=1.01; ▲ denotes the numerical simulation of the BPL model as $q$=1.1; — denotes the theoretical prediction of the BPL model as $q$=1.1; * denotes the numerical simulation of the BPL model as $q$=1.2; — denotes the theoretical prediction of the BPL model as $q$=1.2.

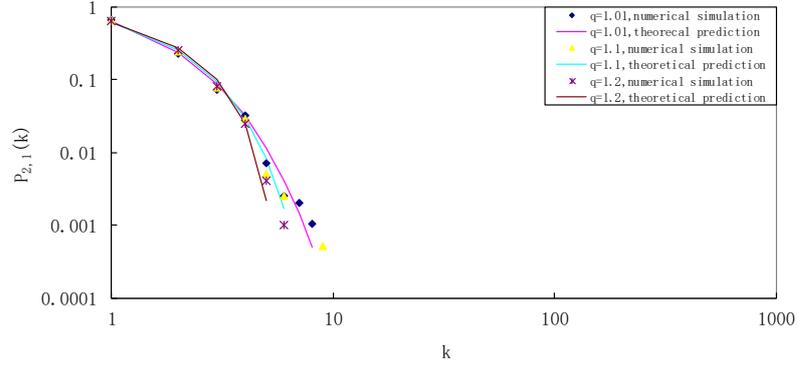

**Fig. 6. The distribution of the 2-order degree of 1-dimensional simplices in the model ( $p = 1 - q < 0$ ).**The figure is in log–log scale. ♦ denotes the numerical simulation of the BPL model as $q$=1.01; — denotes the theoretical prediction of the BPL model as $q$=1.01; ▲ denotes the numerical simulation of the BPL model as $q$=1.1; — denotes the theoretical prediction of the BPL model as $q$=1.1; * denotes the numerical simulation of the BPL model as $q$=1.2; — denotes the theoretical prediction of the BPL model as $q$=1.2.

## 5. Conclusion

This paper proposes the BPL model of higher-order networks based on simplidies, which is a non-homogeneous linear preferential evolution model and is a continuously growing network. The NGF model is a special case of this model as feature parameter $p = -1, 0, 1$.

Analytical results, established here, show that the stationary average degree distribution of the $d$ -order degree of the bending power law of a higher-order network is asymptotically independent of time (and, subsequently, independent of the network size $N = \lambda t + m_0(d+1)$ ). In particular, batch effects of low-dimensional simplex are considered when a new node enters the network. This end indicates that despite its continuous growth, the network reaches a stationary state that is approximately scale-free.

Although the $d$ -order degree distribution of $d-1$ -dimensional simplicies and $d-2$ -dimensional simplicies in the BPL model are influenced by feature parameter $p$ , the $d$-order degree distribution of $e(< d-2)$ -dimensional simplocies shows the phenomenon of heavy-tails. Therefore, the BPL model in higher-order networks plays a role similar to that of the BA model in complex networks. That reflects the emergence of bending scaling in higher-order networks. However, the study on the structure and dynamics of higher-order networks has just begun. It turns out to be more complex than that of complex networks. There is much work to be done on the evolutionary structure of higher-order networks.

**Acknowledgments**

The authors acknowledge the supports of National Natural Science Foundation of China (Grant no. 71971139, 71571119).